\documentclass[aps,twocolumn,prd,epsf]{revtex4}
\usepackage{amssymb}
\usepackage{graphicx}
\newcommand{\be}{\begin{equation}}
\newcommand{\ee}{\end{equation}}
\newcommand{\bea}{\begin{eqnarray}}
\newcommand{\eea}{\end{eqnarray}}

\newcommand{\lpl}{\l_{\rm Pl}}

\begin{document}
%\begin{titlepage}

%\begin{titlepage}

%\vspace{1in}

%____________________________________________________________________________
\title{Oscillatory Universes in Loop Quantum Cosmology and Initial
Conditions for Inflation}
% in Loop quantum Gravity}
%____________________________________________________________________________
\author{James E. Lidsey, David J. Mulryne, N. J. Nunes and Reza Tavakol}

\affiliation{Astronomy Unit, School of Mathematical Sciences, 
Queen Mary, University of London,  
London E1 4NS, U.K.}
%____________________________________________________________________________
%\date{May 2004}
%____________________________________________________________________________
\begin{abstract}
%____________________________________________________________________________
Positively-curved, oscillatory universes are studied within 
the context of Loop Quantum Cosmology
subject to a consistent semi-classical treatment.
The semi-classical effects are reformulated in terms of an effective 
phantom fluid with a variable equation of state. In cosmologies 
sourced by a massless scalar field, 
these effects lead to a universe that undergoes ever-repeating 
cycles of expansion and contraction. 
The presence of a self-interaction potential for the field breaks the 
symmetry of the cycles and can enable the oscillations to establish 
the initial conditions for successful slow-roll inflation, 
even when the field is initially at the minimum of its potential with a 
small kinetic energy. 
The displacement of the field from its minimum
is enhanced for lower and more natural values of 
the parameter that sets the effective quantum gravity scale. For sufficiently 
small values of this parameter, 
the universe can enter a stage of eternal self-reproduction. 

%______________________________
\end{abstract}
%______________________________
%\pacs{98.80.Cq,04.60.Pp}

\maketitle

%______________________________
\section{Introduction}
\setcounter{equation}{0}
%______________________________

Cyclic/oscillatory universes have a long history in cosmology
\cite{Tolam}. Originally, one of their main attractions was 
that initial conditions could in principle be avoided.
Closer consideration of such models, however, revealed severe
difficulties in their construction within
the context of general relativity (GR).
Apart from entropy constraints, which
restrict the number of bounces in the past,
the central difficulty is that when treated
classically any bounce would be singular, thereby resulting in the
breakdown of GR. Recent developments in M-theory 
inspired braneworld models have renewed
interest in cyclic/oscillatory universes \cite{cyclic}, although
problems still remain in such models in developing   
a successful treatment of the bounce. 
An oscillating universe that ultimately undergoes inflationary expansion
after a finite number of cycles has also 
been investigated \cite{kanekar-etal2001}. However, 
a physical mechanism for inducing the bounces was not employed
in this model.  

Our aim here is to study oscillatory universes
within the context of Loop Quantum Cosmology (LQC)
which is the application of Loop Quantum Gravity (LQG) 
to an homogeneous minisuperspace environment.
LQG is at present the main
background independent and non--perturbative
candidate for a quantum theory of gravity (see for example
\cite{rovelli98,thiemann02}).
This approach provides a (discrete) description of high--energy dynamics 
in the form of a difference equation.
An important consequence of this discretization
is the removal of the initial singularity \cite{martin1}.
As the universe expands and its volume increases,
it enters an intermediate semi--classical phase in which the
evolution equations take a continuous
form but include modifications due to non--perturbative
quantization effects \cite{martin3}.
It has recently been shown that a 
collapsing positively--curved Friedmann-Robertson-Walker (FRW) 
universe containing a massive scalar field can undergo a non--singular bounce 
within the semi-classical region \cite{st03}. 
Other studies have shown that for a flat universe 
semi-classical LQC effects increase
the parameter space of initial conditions for successful 
inflation, and this behavior is found to be robust to ambiguities in the theory 
\cite{tsm,blmst}. 

We study the behaviour of a positively--curved universe sourced by 
a massless scalar field and show that the combined effects of the LQC
corrections and the curvature enable the universe to
undergo ever-repeating cycles. 
The presence of a self--interaction potential for the field 
breaks the symmetry of the cycles
and can establish the conditions for inflation,
even if the field is
initially located in the minimum of its potential. 
The viability of this mechanism is studied by identifying 
the constraints that need to be satisfied  
for a consistent semi-classical treatment.

%______________________________
\section{Effective Field Equations in Loop Quantum Cosmology}
%______________________________

We consider positively--curved FRW 
cosmologies sourced by a scalar field $\phi$ with self--interaction 
potential $V(\phi )$. 
The semi--classical phase of LQC arises when the scale factor 
lies in the range $a_i < a < a_*$, where $a_i \equiv \sqrt{\gamma}\lpl$, 
$a_*  \equiv \sqrt{\gamma j/3}  \lpl$, 
$\gamma = \ln 2/\sqrt{3}\pi \approx 0.13$ and 
$j$ is a quantization parameter which must 
take half integer values. Below the scale $a_i$, the discrete nature of 
spacetime is important, whereas the standard classical cosmology is recovered
above $a_*$. The parameter $j$ therefore sets the 
effective quantum gravity scale. 
The modified Friedmann equation is given by 
\be \label{friedeq}
H^2 \equiv \left(\frac{\dot{a}}{a}\right)^2 = 
\frac{8\pi\lpl^2}{3} \, \left[ \frac{1}{2} \frac{\dot{\phi}^2}{D}  +
V(\phi)\right] -\frac{1}{a^2}\, ~,
\ee
where the quantum correction factor $D(q)$ is defined by 
\begin{eqnarray}
\label{defeigenvalue}
D(q) &=& \left(\frac{8}{77}\right)^6 q^{3/2} 
\left\{ 7\left[(q+1)^{11/4} - |q-1|^{11/4}\right]
\right. \nonumber \\
&-& \left. 11q\left[(q+1)^{7/4}- {\rm sgn} (q-1)|q-1|^{7/4} 
\right]\right\}^6 \,
\end{eqnarray}
with $q \equiv (a/a_*)^2$. Eq. 
(\ref{defeigenvalue}) represents an approximate expression for the 
eigenvalues of the inverse volume operator \cite{martin3}. 
As the universe evolves through
the semi--classical phase, this function 
varies as $D \propto a^{15}$ for $a \ll a_*$, has a global 
maximum at $a \approx a_*$ and falls monotonically to $D = 1$ 
for $a > a_*$.  

The scalar field equation has the form
\be
\label{fieldeq}
\ddot{\phi} = - 3 H \left( 1- \frac{1}{3} \frac{d \ln D}{d \ln a} \right)
\dot{\phi} - D V' ,
\ee
where a prime denotes differentiation with respect to the scalar field.
Differentiating Eq. (\ref{friedeq}) and substituting Eq. (\ref{fieldeq}) 
then implies that   
\be
\dot{H} =  -\frac{4\pi \lpl^2 \dot{\phi}^2}{D}
\left( 1-\frac{1}{6} \frac{d \ln D}{d \ln a}  \right) +\frac{1}{a^2} .
\label{raycheq}
\ee

Equations (\ref{friedeq}) and (\ref{fieldeq}) can be written in the
standard form of the Einstein field equations sourced by a perfect fluid: 
\begin{eqnarray}
\label{standard1}
H^2 &=& \frac{8 \pi \lpl^2}{3} \rho_{\rm eff} - \frac{1}{a^2} \,, \\
\label{standard2}
\dot{\rho}_{\rm eff} &=& - 3H \left( \rho_{\rm eff} + p_{\rm eff}
\right) \,,
\end{eqnarray}
where 
\begin{eqnarray}
\label{densityf}
\rho_{\rm eff} &\equiv& \frac{1}{2} \frac{\dot{\phi}^2}{D} + V \,, \\
\label{pressuref}
p_{\rm eff} &\equiv& \frac{1}{2} \frac{\dot{\phi}^2}{D} 
\left( 1- \frac{1}{3}   \frac{d \ln D}{d \ln a} 
\right) -V \,,
\end{eqnarray}
define the effective energy density and pressure of the fluid, 
respectively. 
The effective equation of state, $w \equiv p_{\rm eff} 
/\rho_{\rm eff}$, is given by
\begin{equation}
\label{wf}
w =-1 + \frac{2 \dot{\phi}^2}{\dot{\phi}^2 +2DV} 
\left( 1- \frac{1}{6} \frac{d \ln D}{d \ln a} \right)  .
\end{equation}

In this picture the effects of the LQC corrections on the cosmic dynamics 
are parametrized entirely in terms of the equation of state. 
When $d \ln D /d \ln a >6$, the fluid represents `phantom' matter $(w < -1)$ 
that violates the null energy condition 
$(\rho_{\rm eff} + p_{\rm eff} \ge 0)$ and,  
numerically, this is equivalent to $a < a_{\rm ph} \equiv 0.914a_*$. 
This condition depends only on the quantization parameter $j$ and  
is independent of the potential. Since, within the context of GR, 
a violation of the null energy condition can lead to a non--singular bounce,
this provides an alternative description of how LQC effects can result 
in a bouncing cosmology when the   
universe contracts below a critical size: 
the decreasing energy density of the phantom fluid
during the collapsing phase is eventually balanced
by the growing curvature term in the Friedmann equation.

In the following Sections, we develop this picture further 
for the case of massless and self--interacting scalar fields, respectively.

%______________________________
%\section{Oscillatory Universes in Loop Quantum Cosmology}
%______________________________

\section{Oscillations with a Massless Scalar Field}

\begin{figure}[!t]
\includegraphics[width=8.5cm]{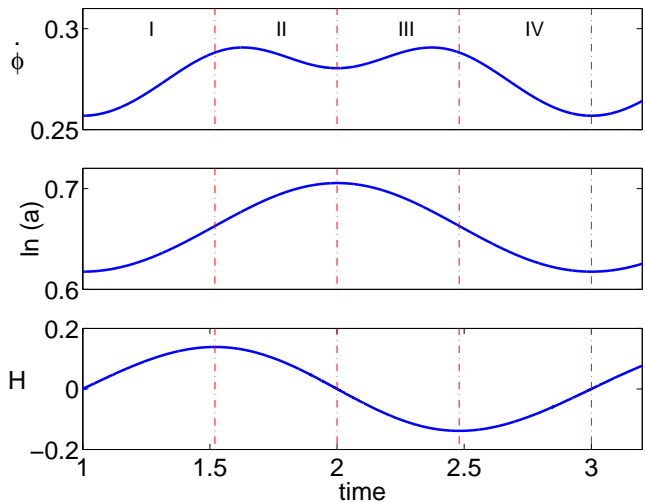}
\caption[] {\label{fig1}
Time evolution of the scalar field $\phi$, the logarithmic 
scale factor and the Hubble parameter when the potential 
$V=0$, 
$j = 100$, $a_{\rm init}/a_* = 0.9$ and $H_i=0$. Axes are labeled in Planck
units.}

\end{figure}

The field equation (\ref{fieldeq}) for a massless scalar field 
$(V=0)$ admits the first integral:
\begin{equation}
\label{firstintegral}
\dot{\phi} = \dot{\phi}_{\rm init} \left( 
\frac{a_{\rm init}}{a} \right)^{3}
\left( \frac{D}{D_{\rm init}} \right) \,,
\end{equation}
where $D_{\rm init} = D(a_{\rm init}/a_*)$ and a subscript 
`init' denotes initial values. Without loss of generality 
we choose initial conditions 
such that $a_i < a_{\rm init} <a_{\rm ph}$, 
$\dot{\phi}_{\rm init} >0$ 
and $H_{\rm init} = 0$. 
The subsequent cosmic dynamics can then be divided into four  
phases (Fig.~\ref{fig1}).  

{\it Phase I:} 
The universe is effectively sourced by phantom matter since $a< a_{\rm ph}$. 
This drives an epoch of superinflationary expansion $(\dot{H} >0)$ until 
$d \ln D/d \ln a =6$. The scalar field accelerates during this phase.

{\it Phase II:} The expansion rate now slows down.  
The scalar field continues to accelerate until 
the logarithmic slope of the quantum correction function has fallen to  
$d \ln D/ d \ln a = 3$. For $a>a_*$, 
the energy density of the field, $\rho_{\phi} \propto 1/a^6$, 
falls more rapidly than the curvature and the expansion 
eventually reaches a turnaround.

{\it Phase III:} 
Eq. (\ref{fieldeq}) implies that the scalar field begins to accelerate 
immediately after the turnaround and continues to do so  
until the universe has collapsed to the point where $d \ln D/ d \ln a = 3$ once 
more. Shortly afterwards, 
the universe contracts below the critical scale $a_{\rm ph}$, 
at which point the Hubble parameter begins to increase.  

{\it Phase IV:} 
Substituting Eq. (\ref{firstintegral}) into 
Eq. (\ref{densityf}) implies that the effective energy density varies as  
$\rho_{\rm eff} = \dot{\phi}^2/(2D) \propto D/a^6$. It 
is therefore decreasing during this phase, since  
$d \ln D /d \ln a > 6$, whereas the curvature is growing. Hence, 
a scale is reached where the Hubble parameter vanishes instantaneously. 
Eqs. (\ref{friedeq}) and (\ref{raycheq}) imply that the second 
time--derivative of the scale factor is positive during this phase and 
the universe therefore undergoes a non--singular bounce 
into a new phase of superinflationary expansion (Phase I). 

%____________________________________________________________________
%_____________________________________________________________________

\begin{figure}[!t]
\includegraphics[width=8.5cm]{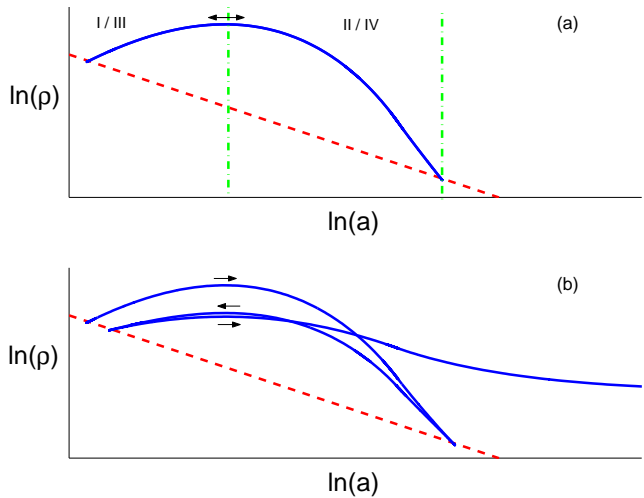}
\caption[] {\label{fig2}
(a) Schematically illustrating the logarithmic 
variation of the effective energy 
density of a massless scalar field (solid line) and the curvature term 
in the Friedmann equation
(dashed line). The universe oscillates indefinitely between the 
intersection points of the two lines. (b)  
Schematically illustrating the effects of introducing a self--interaction 
potential for the field. The cycles are eventually 
broken as the potential becomes dynamically significant, 
thereby resulting in slow--roll inflation.
In both figures, the slope of the trajectories 
is given by $d \ln \rho_{\rm eff} /d \ln a = -3[1+w(a)]$.
}

\end{figure}

The effective equation of state 
(\ref{wf}) is independent of the field's kinetic energy 
when $V=0$. This implies that identical cycles are repeated 
indefinitely into the future (and the past), as summarized  
in Fig.~\ref{fig2}a. During the collapsing phases, the field 
retraces the trajectory it mapped out during the expanding phases.
Oscillatory behaviour is therefore possible 
in the absence of an interaction potential. Furthermore, 
Eq. (\ref{firstintegral}) implies that the field's kinetic energy 
never vanishes during the cycle since 
the scale factor remains finite. The value of the field 
therefore increases monotonically with time.

%____________________________________________________________________
%_____________________________________________________________________

\section{Self-Interacting Scalar Field}

We now consider how self--interactions of the 
scalar field modify this cyclic dynamics. 
We make very weak assumptions 
about the potential, specifying only that it 
has a global minimum 
at $V_{\rm min} (0)=0$ and is a positive--definite and monotonically 
varying function when $\phi \ne 0$ such that $V''>0$. We suppose 
the field is initially located at the minimum of its potential
with the same initial conditions as those considered in Section III. 

The qualitative dynamics of the universe
is illustrated in Fig. \ref{fig2}b.  
In general, the path of the field in the $\{ \ln \rho_{\rm eff}, \ln a \}$
plane is determined by the variation 
of the effective equation of state (\ref{wf}). 
The gradient of the trajectory 
is given by $d \ln \rho_{\rm eff} /d \ln a = -3[1+w(a)]$ and 
there is a turning point whenever the scale factor passes 
through $a_{\rm ph}$, regardless of the 
form of the potential. 

Two factors determine how 
the equation of state changes during each cycle
and these can be parametrized by defining the quantities 
${\cal{B}} = [(6-d\ln D/d \ln a)/6]$ and 
${\cal{C}} \equiv 2\dot{\phi}^2/
[\dot{\phi}^2 +2DV]$, respectively. 
For a massless field, 
${\cal{C}} =2$, and the deviation of the equation of
state away from the value 
$w =-1$ is determined by the magnitude of ${\cal{B}}$. 
The effects of the potential are parametrized by ${\cal{C}}$.
Introducing a potential energy into the system 
necessarily implies ${\cal{C}} <2$ and it follows that for a given 
value of the scale factor, the equation of state is closer to $-1$
than in the massless case.  

Since the field moves monotonically up the potential, 
the gradient term $DV'$ in the scalar field equation (\ref{fieldeq}) 
is more significant in a given cycle relative to the 
previous one. This implies that the scalar field gains kinetic 
energy less rapidly 
during Phases I and III and loses kinetic energy more rapidly during 
Phases II and IV. Consequently, the value of $C$ at the 
end of Phase I of a given cycle is smaller than the value it had at 
the corresponding point of the previous cycle. 
In other words, the rate of change of the 
trajectory's gradient around $a_{\rm ph}$ becomes progressively 
smaller with each successive cycle. 

A massless field begins 
accelerating immediately after the turnaround and bounce have been attained. 
However, Eq. (\ref{fieldeq}) implies that 
$\ddot{\phi} = - DV' <0$ at the instant when the Hubble 
parameter vanishes, so the field does not accelerate immediately 
after the turnaround (bounce);  
there is a short delay during which ${\cal{C}}$ 
continues to decrease. The trajectory of the field
immediately before the turnaround (bounce) is therefore slightly steeper 
than the trajectory immediately after, i.e., 
the trajectory for Phase III (I) lies {\em below} that of Phase II (IV).
Moreover, because the field does not acquire as much kinetic energy 
during Phase III as it lost during Phase II, 
the value of ${\cal{C}}$ (for a 
given value of the scale factor) is smaller 
during Phase III than it was during Phase II. Thus, the
Phase III part of the 
trajectory always lies below that of Phase II and,  similarly, 
the trajectory of Phase I lies below that of the Phase IV 
trajectory of the previous cycle. The potential therefore breaks the 
symmetric cycles of the massless field
and the effective energy density at the end of Phases I and III falls 
with successive cycles.

%The potential therefore breaks the 
%symmetric cycles of the massless field. The value of the
%effective energy density at the end of Phases I and III falls with successive 
%cycles and the rate at which the equation of state moves away 
%from $-1$ at the start of Phase II is slower for each subsequent cycle. 

\begin{figure}[!t]
\includegraphics[width=8.5cm]{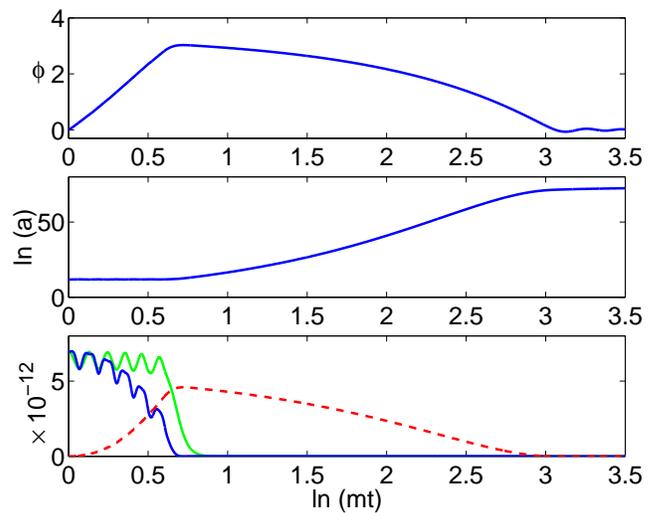}
\caption[] {\label{fig3}
Illustrating the time evolution of the scalar field $\phi$ (top panel) 
and the logarithmic scale factor (middle panel) for a quadratic 
potential $V=m^2\phi^2/2$. The bottom panel 
illustrates how the effective kinetic energy $\dot{\phi}^2/2D$
(solid line) and 
potential energy (dashed line) 
of the field vary compared to the curvature term (grey line).  
The cycles end and slow-roll inflation commences
at the point when the potential begins to dominate the kinetic energy.
Initial conditions are chosen such that $\phi_{\rm init} =
H_{\rm init} =0$ and $a_{\rm init}/a_* = 0.9$, with 
$m = 10^{-6} \lpl^{-1}$ and $j = 5 \times 10^{11}$ 
(see the text for details). The axes are labeled in Planck units. 
}

\end{figure}

We have implicitly assumed that
the kinetic energy of the field never vanishes during Phase II. 
However, the potential becomes progressively more important with 
each completed cycle and  
this assumption must necessarily break down after a finite number of 
cycles have been completed.  Since the effects of the quantum correction 
function are negligible once the universe has expanded beyond 
$a_*$,  the energy density of the field during Phase II
redshifts more rapidly than the curvature term if  
the strong energy condition is satisfied, i.e., 
if $\dot{\phi}^2  >V$. As the field moves monotonically up the 
potential, it becomes progressively harder to maintain this  
condition. Eventually, therefore, a cycle is reached where this
condition is violated during Phase II. This leads
inevitably to an epoch of slow--roll 
inflationary expansion, as the field slows down, 
reaches a point of maximum displacement and moves back down the potential
\footnote{For some initial conditions and choices of parameters, 
this may occur during the first cycle. In this case, 
the scenario discussed in Refs.~\cite{tsm,blmst} is recovered and we do 
not consider this possibility further.}. 
The transition into a potential--driven, slow--roll 
inflationary epoch is shown in Figs. \ref{fig2}b and \ref{fig3}.  
 
Thus far, we have discussed the cyclic dynamics for a given value of the 
quantization parameter, $j$. In spatially flat models, 
the maximum value attained by the field as it moves up the 
potential increases for higher values of $j$, since the universe 
undergoes greater expansion before 
the end of the superinflationary, semi--classical phase.
For positively--curved models, on the other hand, the 
field moves further up the potential before the onset of 
slow--roll inflation for {\em smaller} values of $j$. 
This behaviour can be understood qualitatively 
by making some simplifying assumptions. 
Let us assume 
that the transition from Phase I to II occurs 
at $a = a_*$; that  
$D \propto a^{15}$ for $a< a_*$; that $D=1$ for $a>a_*$; and 
that the potential 
becomes dynamically significant only during the last 
cycle before slow--roll inflation. At this 
level of approximation, the field's kinetic
energy varies as $\dot{\phi}^2 |_{a<a_*} \propto a^{24}$ and 
$\dot{\phi}^2 |_{a>a_*} \propto a^{-6}$, respectively, so 
$\dot{\phi}^2_* \propto a^{24}_* \propto j^{12}$ at the transition.  
At the turnaround $(a=a_t)$, 
$\dot{\phi}^2_t = \dot{\phi}^2_* (a_*/a_t)^6$, but since 
$\dot{\phi}^2_t \approx 1/a^2_t$, it follows that
$\dot{\phi}^2_t \propto j^{-15/2}$. Hence,  
the kinetic energy of the field at the turnaround is higher 
for lower values of $j$ and, consequently, slow-roll inflation occurs
($V>\dot{\phi}^2$) when the field is further from the minimum of the potential. 

To summarize, the cyclic nature of positively--curved
cosmologies within 
LQC can in principle establish the conditions for slow--roll inflation even
if the inflaton is initially located in the minimum of its potential.
We now proceed to consider the parameter space of a viable model of 
inflation.

\section{Parameter space of a viable model}
 
A number of constraints must be satisfied by any successful 
inflationary scenario of the type outlined above. 
In particular, the field must be 
sufficiently displaced from the minimum of its potential at the end of the 
oscillatory phase for the horizon problem to be solved. Typically this 
requires at least $60$ e--folds of accelerated expansion. 
Self--consistency of the semi--classical analysis also requires that:
(a) $H^2$ must be non--negative;
(b) the Hubble length must be larger
than the limiting value of the scale factor, $|H|a_i  < 1$; and
(c) the scale factor at the bounce must exceed $a_i$. Imposing constraint (b) 
is equivalent to requiring that energy scales during the 
classical regime do not exceed the Planck scale \cite{blmst}.

%____________________________________________________________________
\begin{figure}
\includegraphics[width=8.5cm]{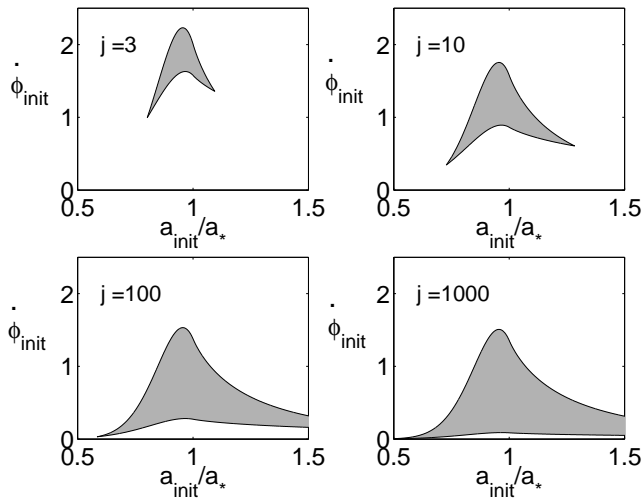}
\caption[] {\label{fig4}
Constraints on the initial velocity of the field $|\dot{\phi}|_{\rm 
init}$ for a
given initial value of the scale factor, 
$a_{\rm init}/a_*$, for different values of the quantization 
parameter, $j$. The axes are labeled 
in Planck units and the shaded areas represent the regions where
constraints (\ref{phidotmin}) and (\ref{phidotmax}) are satisfied. 
For all values of $j$, the area of the shaded region is 
finite, and the points of intersection occur further from 
$a_{\rm init}/a_* \approx 1$ as $j$ is increased. Note that for 
$a_{\rm init} > a_*$, the universe is initially in a contracting phase and
subsequently 
bounces, after which the behaviour discussed in the text is followed.
}
\end{figure}
%_____________________________________________________________________
%

Constraint (a) necessarily implies that 
the initial kinetic energy of the field is bounded 
from below by the Friedmann equation (\ref{friedeq}):
\be
\label{phidotmin}
\dot{\phi}_{\rm init}^2 >  
\frac{6D_{\rm init}}{8\pi\lpl^2a_{\rm init}^2} \,.
\ee
On the other hand, constraint (b) results in an upper bound on 
the field's kinetic energy. Eq. (\ref{raycheq}) 
implies that $|H|$ is maximized when $d\ln D/d \ln a \approx 6$ and 
substituting Eq. (\ref{firstintegral}) into the Friedmann 
equation (\ref{friedeq}) then implies that $|H_{\rm max} | a_i <1$ for
\begin{equation}
\label{phidotmax}
\dot{\phi}_{\rm init}^2 < \frac{3}{4\pi \lpl^2} 
\left(\frac{a_{\rm ph}}{a_*}\right)^6 
\left(\frac{a_*}{a_{\rm init}}\right)^6
\frac{D_{\rm init}^2}{D_{\rm ph}}
\left(\frac{1}{\gamma\lpl^2}+\frac{1}{a_{\rm ph}^2}\right) \,,
\end{equation}
where $D_{\rm ph} = D(a_{\rm ph}/a_*)$. The discussion leading to 
Fig. \ref{fig2}b has shown that the maximum value of 
the square of the Hubble parameter during a given cycle is smaller 
than that of the previous cycle. Thus, if condition (\ref{phidotmax}) 
is satisfied in the first cycle, where the potential is negligible, 
$|H|a_i <1$ during all subsequent cycles.

Fig.~\ref{fig4} illustrates 
the region of parameter space where constraints 
(\ref{phidotmin}) and (\ref{phidotmax}) are simultaneously satisfied.
These constraints are independent of the potential. 
Increasing the value of $j$ reduces the upper and lower
bounds on the initial kinetic energy but widens the range of allowed 
values of $a_{\rm init}/a_*$. 
A further consequence of Fig. \ref{fig4} is that for a given value of $j$
there exists an upper limit to the size of the universe at the 
turnaround that is 
consistent with the semi--classical dynamics. In other words, 
if the universe is too large at the turnaround, 
the kinetic energy of the scalar 
field will exceed the Planck scale before the universe has contracted below 
$a_*$. This implies, in particular, that 
if the inflaton did not decay at the end of inflation, 
future cycles after the slow--roll inflationary epoch could not 
be realized within this semi--classical framework. 

Constraint (c) is not necessarily satisfied 
if the bounce occurs at progressively smaller 
values of the scale factor. It is necessary, therefore, to consider this
constraint for each specific model. 
We have considered a quadratic potential 
$V(\phi) = m^2 \phi^2/2$, where the mass of the field is set by the COBE 
normalization to be $m=10^{-6}\lpl^{-1}$. 
We have verified numerically that constraints  (a)--(c) remain satisfied for 
all initial values of the parameters contained within the shaded 
regions of Fig. \ref{fig4} and, furthermore, that  
the maximum value attained by the field 
is sufficient to solve the horizon problem, 
i.e., $\phi_{\rm max} > 
3 \lpl^{-1}$. Indeed, for the initial
condition $a_{\rm init} = 0.9 a_*$ considered in Fig. \ref{fig3}, 
sufficient inflation is possible for $j \le 5 \times 10^{11}$. 
Numerical integration indicates that successful slow--roll inflation 
is also possible for other potentials such as quartic and hyperbolic 
models. 

\section{Conclusion}

We have shown that loop quantum gravity corrections to the Friedmann
equation in a positively--curved universe enable a scalar field to move 
up its potential due to a series of contracting and expanding phases 
in the cosmic dynamics. 
The oscillatory nature of such models can be described by 
reformulating the semi--classical dynamics in terms of an 
effective phantom fluid. In general, the field is able to move further 
from the potential minimum for lower and more natural
values of the quantization ambiguity parameter $j$. 
This is important because for sufficiently small $j$, the field may move 
far enough up the potential before the cycles 
are broken for the universe to enter a stage of eternal
self--reproduction \cite{linde}. Eternal inflation occurs if 
$3 V'^2 < 128 \pi \lpl V^3$ and for a quadratic potential 
this implies that $\phi > 1/2 m^{1/2} \lpl^3 
\approx 500 \lpl^{-1}$ \cite{linde}. Numerical simulations 
indicate, for example, 
that this condition is satisfied for $j < 2 \times 10^7$ when 
$a_{\rm init} = 0.9 a_*$. 
We conclude, therefore, that 
even if the field is at the minimum of its
potential, a wide range of initial conditions 
leads to successful (and eternal) inflation. 
Since the assumptions made about the form of the potential were weak,
this mechanism should be very generic.

Finally, the oscillatory dynamics also arises when the minimum of the potential 
is at $V_{\rm min} <0$. 
In positively--curved universes, the same mechanism enables the 
field to work its way out of the negative region and 
continue up the potential until the cycles are broken, as discussed in 
Section IV. For spatially flat universes, the picture is similar
although the oscillations come to end as soon as the 
field reaches a positive region of the potential. This is particularly
interesting given that negative potentials are known to arise in
string/M-theory compactifications (e.g. \cite{Emparan:2003gg}). 
It would be interesting to explore these possibilities further. 

%__________________________________________________

%_____________________________________________________________________________
%_____________________________________________________________________________

\begin{thebibliography}{tbds}
%__________________________________________________

\bibitem{Tolam} 
R. C. Tolman, {\em Relativity, Thermodynamics and Cosmology}
(Clarendon Press, Oxford, 1934).

\bibitem{cyclic}  
J. Khoury, B. A. Ovrut, P. J. Steinhardt and N. Turok, 
Phys. Rev. D {\bf 64}, 123522 (2001) [arXiv:hep-th/0103239]; 
J. Khoury, B. A. Ovrut, N. Seiberg, P. J. Steinhardt and N. Turok,
Phys.\ Rev.\ D {\bf 65}, 086007 (2002) [arXiv:hep-th/0108187];
P. J. Steinhardt and N. Turok,
Science, {\bf 296}, 1436 (2002); Phys.\ Rev.\ D {\bf 65}, 126003 (2002)
[arXiv:hep-th/0111098]; arXiv:astro-ph/0404480.

\bibitem{kanekar-etal2001} 
N. Kanekar, V. Sahni and Y. Shtanov, 
Phys.\ Rev.\ D {\bf 63}, 083520 (2001) [arXiv:astro-ph/0101448].
 
\bibitem{rovelli98}
C. Rovelli, Liv. Rev. Rel. {\bf 1}, 1 (1998) [arXiv:gr-qc/9710008].

\bibitem{thiemann02}
T.~Thiemann, Lect.\ Notes Phys.\  {\bf 631}, 41 (2003) [arXiv:gr-qc/0210094].

\bibitem{martin1}
M.~Bojowald, Phys.\ Rev.\ Lett.\  {\bf 86}, 5227 (2001)
[arXiv:gr-qc/0102069]; 
Phys.\ Rev.\ Lett.\  {\bf 87}, 121301 (2001) [arXiv:gr-qc/0104072].

\bibitem{martin3}
M.~Bojowald, Phys.\ Rev.\ Lett.\  {\bf 89}, 261301 (2002)
[arXiv:gr-qc/0206054].

\bibitem{st03} P.~Singh and  A.~Toporensky, 
Phys. Rev. D {\bf 69}, 104008 (2004) [arXiv:gr-qc/0312110].

\bibitem{blmst} M.~Bojowald, J. E. Lidsey, D. J. Mulryne,
P. Singh and R. Tavakol, In Press, Phys.\ Rev.\ D [arXiv:gr-qc/0403106].

\bibitem{tsm} S.~Tsujikawa, P.~Singh and R.~Maartens,
arXiv:astro-ph/0311015.
 
\bibitem{linde} 
A. D. Linde, Phys. Lett. {\bf 175B}, 395 (1986). 

\bibitem{Emparan:2003gg}
P. K. Townsend and M. N. R. Wohlfarth, Phys. Rev. Lett. {\bf 91}, 
061302 (2003) [arXiv:hep-th/0303097]; 
N. Ohta, Phys. Rev. Lett. {\bf 91}, 061303 (2003) [arXiv:hep-th/0303238]; 
R.~Emparan and J.~Garriga, JHEP {\bf 0305}, 028 (2003) [arXiv:hep-th/0304124]; 
N. Ohta, Prog. Theor. Phys. {\bf 110}, 269 (2003) [arXiv:hep-th/0304172]; 
C.-M. Chen, P.-M. Ho, I. P. Neupane, N.Ohta, and J. E. Wang, 
JHEP {\bf 0310}, 058 (2003) [arXiv:hep-th/0306291]. 




%_____________________________________________________________________________
\end{thebibliography}
\end{document}